\documentclass[aps,showpacs, 12pt]{revtex4}
\usepackage{amsmath, amssymb,color,graphicx,subfigure}
\usepackage{float}
\begin{document}
\draft
\title{Velocity Autocorrelation Function of Magnetized
Two-Dimensional One-Component Plasma}
\author{Girija S. Dubey and  Godfrey Gumbs }
\address{Department of Physics and Astronomy\\
Hunter College of the City University of New York\\
695 Park Avenue, New York, NY 10065}

\begin{abstract}
The velocity autocorrelation function (VAF) for a two-dimensional one-component plasma (OCP)
 is investigated by employing molecular dynamics techniques. The VAF exhibits well defined
oscillations whose frequency is independent of the dimensionless
 coupling parameter $\Gamma$. However,  the presence of a uniform
 perpendicular  external magnetic field increases the height of the first peak.
Molecular dynamics computer simulation results are presented
for a two-dimensional OCP in a perpendicular magnetic field
as a function of the Plasma parameter.  Our simulation
results clearly indicate  that at high temperature, the magnetic
field affects the VAF in a substantial way.
\end{abstract}


\maketitle

\section{Introduction}
\label{sec:1}


A great deal of effort has been devoted to gaining a better understanding of
the optical and electron transport properties of one-component plasmas, which
may be treated as   two-dimensional electron gases (2DEG)  or
three-dimensional electron gases (3DEG) [\onlinecite{RC2015,GOD1,GOD2,GOD3}]. This is due in part
to 2DEG structures being used in models of semiconductor heterostructures. In contrast
to its three-dimensional counterpart, the 2DEG is by far more challenging because
of the quantum effects arising from  the confinement. In studying
static properties, both integral equation theory and Monte Carlo simulations have
been developed. The hypernetted chain (HNC) approximation integral equation theory
has been proved to be quite accurate compared to simulations based on phenomenological models  \cite{GOD2}, but it
requires more elaborate  numerical calculations due to the time consuming Frankel transform,
i.e., the two-dimensional Fourier transform. Moreover, in the absence of a
magnetic field, dynamical properties have been investigated through molecular dynamics
(MD) and Kubo-Greenwood short time expansions. Good agreement between these
two approaches has been reported in the literature \cite{KG1}.

\medskip
\par

Fabrication of high mobility 2DEGs is desirable in the  semiconductor industry.
In these structures, electrons are confined in a layer between two carefully
chosen semiconductor materials, such as GaAs and AlGaAs. These have found
useful applications in semiconductor devices as a result of the unique electron
transport properties at the interface. In low magnetic
fields,  applied perpendicular to the 2DEG, unusual
transport properties have been observed.  Apart from the well-known
magnetoresistance oscillations which are due to commensurability
effects in modulated 2D systems \cite{s1,s2,s3,s4,s5},  a number of other
low magnetic field anomalies appear at low temperature.  For example,  an
observed anomaly is the positive magnetoresistance followed
by a resistance drop at higher magnetic fields.  Such behavior was
observed in  metallic single crystals \cite{s6,s7}. It has been
successfully explained by the suppression of Bragg reflections due to
the Lorentz force induced by the magnetic field, an effect known as
magnetic breakdown \cite{s8}.

\medskip
\par

A considerable amount of information is available for the
electron transport and optical properties of the 2DEG \cite{rr,rr01,rr02} from both an
experimental and theoretical point of view in the absence of a magnetic
field. However, the effect of a perpendicular magnetic field  could lead to
novel behavior when the Coulomb interaction cannot be neglected
 \cite{rr1}. Since an understanding of the physics of these systems  in an
applied magnetic field is helpful in device and metrological applications,
there is much interest in their properties as a function of an applied field.
For example, Cunningham, et al. \cite{rr2} and Wright, et al. \cite{Wright}
have analyzed the effect of a
perpendicular magnetic field on the plateaus of an acoustoelectric current in
a narrow channel within a 2DEG. Their goal was to examine how the quantization
of the acoustoelectric current was affected by magnetic field.

\medskip
\par

There has been a great deal of interest in the static,
dynamic and thermodynamic properties of two-dimensional (2D) electron
systems on Helium and other substrates \cite{s1}. More recently, there has
been much effort regarding how a substrate affects the collective modes of
graphene \cite{Graphene1,Graphene2}. The 2D system
we are concerned with is described by the Coulomb potential which
is the solution of Poisson's equation, and is thus given by
$U(r) = - e^2 \log { (r/L)}$, where $r$ we the distance between
the two charges and $L$ is a scaling length which we choose
as the length of the simulation cell. Such a 2D one-component
plasma (OCP) consists of identical point particles
carrying a charge $-e$ and interacting through the Coulomb potential.
To ensure charge neutrality, the particles are immersed in a uniform
positive background of charge. The thermodynamic state of the OCP
is completely defined by the dimensionless plasma coupling parameter
$\Gamma = e^2/ { k_B T} $, where $a$ is the radius
of the Wigner-Seitz disk and $T$ is the temperature.
Depending on the value of the coupling
parameter, the system can be in the weakly correlated gaseous phase,
strongly correlated liquid phase or in the crystalline Wigner lattice
state. There is a reasonably good understanding theoretically for the
2D electron system \cite{s2,s3,s4,s5,s8} in a magnetic field
making use of various types of approximations to deal with the
many-body effects. In this paper, we use  molecular dynamics (MD)
simulations.
\medskip

Our calculations are for a simplified model of electron layers
such as those on a liquid helium surface. These systems have been
generated by Grimes and Adams  \cite{s6} as well as other related
experiments \cite{s7}, in which the thickness of the layers of the
electrons trapped on the surface of liquid He is much less than
the inter-particle spacing $a$, and we may thus treat the system
as 2D. As a matter of fact, Ma, Girvin, and Rajaraman \cite{s1}
have argued that even though the Coulomb potential in  quasi-2D
systems is really $1/r$, the effective potential is closer to
$\log(r)$ for a layer with high dielectric constant surrounded by
material with much smaller dielectric constant.
The application of a uniform perpendicular
magnetic field does not affect the classical nature of the system as
long as the separation between Landau levels satisfies
$\hbar \omega_c \ll   k_B T$, where $\omega_c$ is the cyclotron
frequency. The number density, $n$ has been varied over
a wide range, i.e., $ 10^5  \leq n \leq 10^{10}$ cm$^{-2}$ with a Fermi
energy of at most $10^{-3}$ K, for which the system may be treated
classically. We have carried out calculations over a range of coupling
parameters for which there is a variation of the velocity auto-correlation
function (VAF) with magnetic field at finite temperature. Yamada and Ferry
\cite {rr1} have also investigated in detail the effect of magnetic field
on some transport properties of 2D electron systems.  Berne \cite{Berne}
has reported molecular dynamics calculations of the VAF of an OCP in a
magnetic field for a limited number of values for the coupling parameter $\Gamma$.
Additionally,  Dzhumagulova et al. \cite{DZH}  have also investigated the feect due to a
perpendicular magnetic field on the velocity autocorrelation
function for a two-dimensional one-component system by employing
a Yukawa potential and $\Gamma=120$.

\medskip

In this paper, we will present a molecular dynamics study of a classical,
interacting 2D electron system in the presence of a uniform magnetic
field for a range of values of temperature, i.e., the coupling parameter
$\Gamma$. Such systems have been used in the study of high-$T_c$
superconductors, where a logarithmic potential is seen to be appropriate
for the vortices within a single layer. We have calculated the
VAF, root mean square displacement and the self correlation correlation
function as functions of magnetic field and plasma coupling parameter.

Hansen et al. \cite{GOD3},
Kalia et al. \cite{Kalia} and
de Leeuw et al. \cite{DEL2} have studied the
VAF in a   one-component  plasma. Specifically, Hansen et al. \cite{GOD3}
did their calculations in $3$ D. Kalia et al.\cite{Kalia} employed a
$1/r$ potential for selected values of the coupling parameter $\Gamma$.
in the absence of magnetic field. de Leeuw et al. \cite{DEL2}   considered
a 2D one-component plasma  using a $\ln(r)$ potential in the absence of magnetic
field. for a coupling parameter $\Gamma=72$.
We have carried out extensive calculations and we  present here
new results, to see the effect of magnetic field.
Agarwal and Pathak \cite{agarwal} reported results
for the VAF of a 2D classical electron fluid using a short-time
expansion in the absence of a magnetic field.

 \medskip
The rest of the paper is organized as follows.   We first present
details of the computational details of the
simulations. Following this, we will present the result for the velocity correlation function
root mean square displacement and single particle correlation function.
We  discuss these results with a summary in Sec. \ref{sec3}.
and highlights of our results.

\section{Method and Model}
\label{sec:2}

The validity of the present calculations is examined by comparing the results
in the absence and presence  of magnetic field. The effect of the magnetic
field is included through a Lorentz force. The advantage  of our method
enables us to watch the system evolving in real time, which in turn offers one
various possibilities to calculate several time correlation functions required
for comparison with experimental results.

\medskip
\par

In 2D, the Coulomb interaction between two particles of charge
$e$ separated by a distance $r$ is given by $V(r)  =  -e^2 \ln {r} /{L}$\ ,
where $L$ is a scaling length chosen as the length of the simulation cell.
Because of the long-range nature of the
Coulomb interaction, we use the method of Ewald summations to take
into account the interaction of an electron with an infinite array of
periodic images of the other electrons and the uniform, positive
background. In computer simulation studies, one usually deals with
systems composed of a few hundred particles in a rectangle of area
$A$ under periodic boundary conditions. The energy of the system,
composed of $N_0$ particles of charge $e$ embedded in a uniform neutralizing
background, is given by

\begin{eqnarray}
U & = & {e^2\over 2} \sum_{n}
 \sum_{i=1}^{N_0}\ ^{\prime} \sum_{j=1}^{N_0}\ ^{\prime}
 \log{|(\vec{r}_{ij}+\vec{n})|}
 \nonumber\\
  &+& {Ne^2\over A} \sum_{\vec{n}}\sum_{j=1}^N  \int_A d\vec{r}\  \log|\vec{r}_j-\vec{r}+\vec{n}|-
{N^2e^2\over A^2} \sum_{\vec{n}}\int_Ad\vec{r} \int_A  d\vec{r}\ ^{\prime}
\log |\vec{r}-\vec{r}\ ^{\prime}+\vec{n}|\ ,
\label{e2}
 \end{eqnarray}
where $ \vec{r}_{ij}$ is the vector between particles $i$ and $j$,
$ \vec{n}$ is a vector whose components are ($ n_{x}L, n_{y} L $) and
$n_x$, $n_y$  are integers.
The first term on the right-hand side of Eq.\ (\ref{e2}) arises from the
interactions between the particles. The sum over $n$ is carried out over
all lattice vectors. The primes on the summation signs indicate
 that, when $n=0$, the terms with $i=j$ have to be excluded.
The second term arises from the interaction of particles with the
background. The integral is to be taken over the rectangular cell
of area $ A=L_x L_y $. The third term represents the interaction of
the background with itself. The sum in Eq.\ (\ref{e2})
is conditionally convergent
and the energy $U$ depends on the shape of the macroscopically large
periodic system. Adding the contributions of different cells of the
system in circular annuli, Eq.\ (\ref{e2}) can be rewritten in the following
form using the Ewald summation

\begin{eqnarray}
U &= & {e^2\over 4}\sum_{n} \sum_{i=1}^N \sum_{j=1}^{N_0}
E_{1}(\alpha^2 (r_{ij}^2+n ^2))
+ {e^2\over 4\pi A}\sum_{k}\ ^{\prime}
\exp\left({\pi^2k^2\over \alpha^2}\right)
\frac{\rho_k\rho_{-k}}{k^2}
\nonumber\\
&-& {Ne^2\over4}[C +\log(\alpha^2A)]-{N^2e^2\pi\over4\alpha^2A}\ ,
\label{e3}
\end{eqnarray}
where the sum over the wave vector $k$ excludes the divergent term $k= 0$,
as denoted by prime, $ \alpha $ is a convergence parameter which controls
the relative weights of the real space and reciprocal space sums,
$ \vec{k}=2 \pi \vec{n} / L $, and $C=0.5772...$ is
Euler's constant. Also, in this notation,
$E_1(x)$ denotes the exponential integral and the
Fourier-component $\rho_k$ of  the electron density is given by:

\begin{equation}
\rho_k =  \sum_{j=1}^{N_0} \exp(2 \pi i \vec{k}\cdot\vec{r_j})\ .
\label{e4}
\end{equation}
From our numerical calculations, we have found that the value of
$\alpha A^{1/2} = 6.5$ gives a satisfactory
convergence rate for each of the two series in Eq.\ (\ref{e3}),
\medskip

\section{Numerical Results}
\label{sec:3}

\medskip
\par

We now report the results of our calculations for molecular dynamics
simulation experiments for the 2D Coulomb system with  magnetic field.
The calculations were carried out  for $256$ particles and five values of
the plasma parameter $\Gamma  =    e^2/  {  k_BT} $. We chose
$\Gamma= 20,\ 36,\ 50,\ 90$ and $130$ and the
range of magnetic field was $0\leq B^\ast\leq 20$ T, in a rectangular cell
with periodic boundary conditions. A predictor-corrector method
involving up to five time derivatives of the positions was employed
to integrate Newton's equation of motion. The total force ${\bf F}_T$
on an electron is the sum of the forces arising from the Coulomb interaction
and the Lorentz force so that $\displaystyle{{\bf F}_T={\bf F}(U) - { e \over c} ( {\bf v}
\times {\bf B} )}$, where ${\bf v}$  is the velocity of the
electron. We used the following dimensionless variables for
length and time: $\ x^\ast = x /a,\ t^\ast = t/ \tau$, where
$\tau$ = $\sqrt{ma^2/e^2}$ and $m$ is the mass of the particle.
The reduced density is $n^\ast= na^2$ and we chose the electron density as
$n=1.477 \times 10^8$/cm$^2$ and five values of temperature ranging from
$T=7$ mK to $50$ mK. The runs for the collection of data presented here for  $256$
electrons extended over $30\times 10^5$ time steps, after an initial
$10^6$ time steps for the system to reach equilibrium.  During our simulations,
the energy of the system remained constant to within one part in $10^5$. The equations
of motion were integrated with a time step of $\delta t = 0.00123 \tau$.

\medskip
\par

For convenience, we have used dimensionless units for space and time
as follows $\ r^\ast = r /a,\ t^\ast = t/ \tau$, where $\tau = 0.1$ (in units
of $\sqrt{m_a^2/e^2}$). and  $\Gamma  =    e^2/  {  k_BT} $. The reduced density
$n^*= na^2$, which corresponds to the value $1.477 \times 10^8$
electron/cm$^2$ and at 20 values of temperature ranging from $0.19$ to $1.0$ K,
corresponding to values of $\Gamma$ equal to $180.0$ and $36.0$, respectively. The runs for
the collection of the data presented here for  $256$ electrons extended over
$15\times 10^4$ time steps, after an initial $10^5$ time steps for the system
to reach equilibrium. We have done one set of calculations for $N_0=400$ particles to
check the dependence on the value of $N_0$ for the system. It was found that this
energy was independent of $N_0$ within the numerical accuracy of the calculations.
During our simulations, the energy of the system remained constant to within 1 part in $10^4$.

\medskip
\par

The velocity autocorrelation function, which is defined as

\begin{equation}
Z(t) = \frac {< V_i (t)\cdot V_i(o)>} { <V_i(o)^2> }.
\label{e5}
\end{equation}
provides us with some insight into single particle dynamics.

In Fig.\ 1, we have presented our results
for the velocity autocorrelation function when there is no external
magnetic field as well as in the presence of a perpendicular magnetic
field $B^\ast$ = $1,\ 2,\ 5, 10$ and $15$ T and when the coupling
constant $\Gamma$ = $20$. In Fig.\  2, the magnetic field is chosen as $1$ T and several values of
$\Gamma$ are used in our calculations.
The results show a series of well defined oscillations, whose period
is independent of the value of the dimensionless
variable $\Gamma$. This behavior is characteristic
for one component charged fluids and has also been observed in the
three dimensional OCP \cite{GOD1} and the classical 2D electron fluid.
A comparison of the results shows
that the frequency of the oscillation is independent of $\Gamma$, but
for a chosen time, $Z(t)$ is decreased as $\Gamma$ is increased.
From Fig.\ 1,  it is clear that the magnetic
field has no effect on the VAF for the smaller values of $\Gamma$.
We have found that the effect is small if the value of $\Gamma$ is
increased. Furthermore, for large values of $\Gamma$, the simulation
results clearly show that in
the presence of a magnetic field, the amplitudes of the oscillations
are slightly changed, but as the time increases, after a few oscillations,
their difference is negligible.

In Fig.\ 3, the root mean square (RMS) displacement is plotted as a function of time for various values
magnetic field $B^\ast$ for one coupling constant $\Gamma=20$. The reason for plotting the RMS  is to clearly see the
effect of magnetic field on the diffusion constant. The simulation results clearly
show that the diffusion constant does not decrease monotonically with
magnetic field. Additionally, the  RMS displacements at each magnetic field are
investigated in order to determine what is the effect due to  magnetic field
on the diffusion,  The mean square displacement is given by:

\begin{equation}
\left <R^2(t) \right> =  \frac{1}{N} \left <\sum_{j=1}^{N} (r_j(t) - r_j(0))^2 \right> \ .
\label{e6}
\end{equation}
here, $r_j(t)$ is the position of the $j^{th}$ electron at  time $t$,
$N$ the total number of electrons and $ < \cdots> $ stands for taking the time average
of a quantity. In Fig. \ 3, we have presented results for the mean square
displacement as a function of time for $B^\ast$=0.0, 1.0 ,2.0 , 5.0 , 10.0 and 15.0 T.
Parabolic increments are applied for all magnetic fields. The slope of
the linear increments decreases with increasing magnetic field.
Since one can obtain  the diffusion constant from the slope of the
curve, it is clear that for each value of $B^\ast$, the  diffusion constant will
be different. Therefore,  magnetic field has some effect on the dynamics of the
system. By making a comparison between  three values of diffusion
constant, it is apparent that by increasing the strength of the magnetic field, the
diffusion constant is changing about $ 5 \% $, we have run the system quite a long time,
and the behavior will not change if one runs the computer program
even longer.

\medskip
\par

In Fig. \ 4,  we have plotted the self correlation function for $B^\ast$= 1.0 for $\Gamma=20$
, which is defined as

\begin{equation}
F_s(q,t) =\frac{1}{N}\sum_{j}  \exp(iq \cdot (r_j(t)-r_j(o))
\label{e7}
\end{equation}
The self correlation function is directly related to the diffusion constant
It is instructive to investigate whether the self correlation function also
show in its behavior, the effect of magnetic field. We have
plotted the self correlation function for various wave vectors for
each of three values of magnetic field. We will make a comparison
in two ways. First we will plot the self correlation function
for each wavevector for three values of magnetic field, then
chosen value  of magnetic field and various    wavevectors.
Our calculations have shown that as the magnetic field is increased, the self
correlation function is quickly reduced.  This is consistant
with the behavior of the mean square plots. In  other words, the magnetic field has some
clearly affects the dynamics of the single-particle motion of the system, as expected.

\medskip
\par

We conclude fron our results for the pair correlation  function that electrons have no
preferred location. Since electrons cannot approach  each other arbitrarily close
due to the strong repulsive Coulomb  force,
they will preferably  move to neighboring unit cells and the applied  magnetic field
has no control in  deternining which cells they would migrate into.
This is the physical reason why magnetic field has virtually no effect on $g(r)$.
\medskip

It is well known that  relaxation times to equilibruim always favor
shorter paths which correspond to smaller diffusion constant.
This is clear from our results in Figs.\  2 and 3 and are consistent with
a semiclassical approach. In our  clssical formalism, electrons
may move   easily between potential minima and will never get trapped in
potential well giving rise to  some diffusion constant. However, in the presence of
magnetic field, the   the diffusion constant may be zero due to trapping in a potential minimum.
However the applied magnetic field does not only change  the electron
dynamics, but also its energy.
In summary,  we have carried out molecular dynamics simulations  for a
Coulomb system in an external magnetic field. The simulation result
clearly show  the static properties are unchanged due to magnetic field.

\section{ Summary and Conclusions}
\label{sec3}

In Fig.\ 1, we have presented our results for the velocity autocorrelation function
when there is no external magnetic field as well as in the presence of a perpendicular magnetic
field $B^\ast=1,\ 2,\ 5, 10$ and $15$ T and when the coupling constant $\Gamma=20$. In Fig.\ 2,
the magnetic field is chosen as $1$ T and several values of $\Gamma$ are used in our
calculations. The results show a series of well-defined oscillations, whose period
is independent of the value of the dimensionless  variable $\Gamma$. This behavior is a
characteristic for one component charged fluids and has also been observed in the 3D
 OCP \cite{rGGG} and the classical 2D electron fluid.  A comparison of our results shows
that the frequency of the oscillation is independent of $\Gamma$, but
for a chosen time, $Z(t)$ is decreased as $\Gamma$ is increased.
From Fig.\ 1, it is clear that the magnetic field has no effect on the VAF for the
smaller values of $\Gamma$. We have found that the consequence of the applied magnetic field
is not significant if the value of $\Gamma$ is increased. Furthermore, for large values of $\Gamma$,
the simulation results clearly show that in the presence of a magnetic field, the amplitudes
of the oscillations are slightly changed, but as the time increases, after a few oscillations,
their difference is negligible.

\medskip
\par

Figs.\ 1 and 2 provide information about the dynamics of electrons  over intervals of
$\sim 10^{-13}$ sec which can be comparable with the time between particle collisions.
This means that we are probing features of dynamical behavior mainly
determined by electron-electron interactions.

\medskip
\par

The short-time behavior of the velocity autocorrelation function
can be studied with the use of a Taylor series expansion,  i.e.,

\begin{equation}
{\bf v}_i(t)={\bf v}_i(0)+ {\bf v}_i'(0) t + \frac{1}{2} {\bf v}_i^{\prime\prime}(0)
t^2 + \cdots\ .
\label{e6}
\end{equation}
Multiplying by ${\bf v}_i(0)$ and taking an ensemble average, we obtain

\begin{eqnarray}
\langle{\bf v}_i(t) \cdot {\bf v}_i(0)\rangle &=& \langle v_i^2\rangle -
\frac{1}{2} \langle v_i'^2\rangle t^2
+ \cdots
\nonumber \\
&=& \langle v_i^2\rangle \left(1 - \frac{1}{2} \Omega_0^2 t^2 + \cdots\right)\ ,
\label{e7}
\end{eqnarray}
where
$\Omega_0^2=\langle |F|^2\rangle /2mk_B T$. Time reversal symmetry ensures
that the odd terms in $t$ vanish. The short-time behavior of the velocity
autocorrelation function is related to the mean square acceleration
(or mean square force $<|F^2|>$) and provides information about the
interaction between the particles as well as the lattice structure.

\medskip
\par

The second term in the Taylor series expansion of the
VAF arises from the combined effect due to the random force $F$ and the
single-particle velocity  (see Eq.\ (\ref{e7})). Our findings confirm the rapid decay
and oscillatory behavior in the VAF due to  a single binary collision between the
tagged particle and the other particles. Also, the rapid oscillations indicate the
strong coupling with the collective plasma excitations \cite{agarwal}.
The Einstein frequency $\Omega_0$ is for small oscillations in the potential well formed from
the electron-electron interaction. Since our calculations show that the
VAF at intermediate times strongly depends on the electron coupling parameter
$\Gamma$, the electron-electron interaction clearly dominates over the
Lorentz force and at small times, their combined effect is given by the
second term in  Eq.\ (\ref{e7})). As $\Gamma$
is increased, the system becomes disordered due to the random force
field. The correlations between the random force and the velocity
which together produce the oscillations in the VAF
become less important as $\Gamma$ is increased and after a
long time.

\medskip
\par

The Lorentz force on an electron causes it to move on a circular orbit
between collisions. After each collision, which is assumed instantaneous,
the electron again moves on a circular path with a different center but
whose cyclotron radius is unchanged since the energy is conserved.
Consequently, the motion of the electron is equivalent to that
in a randomly applied time-dependent magnetic field
that changes the center of the cyclotron orbit. This explains why
the VAF, a dynamical quantity,  depends on the fixed external
magnetic field. Also, the mean force which is determined by
correlations through the pair correlation function is affeted
by the value chosen for $\Gamma$.

In summary,  we have carried out molecular dynamics simulations
for the VAF of a 2D OCP system in an external magnetic field.
We presented numerical results for a range of values of the
coupling parameter $\Gamma$ and the magnetic field $B^\ast$.
Our simulation results show how the period and amplitude of the
oscillations of the VAF depend on these two variables;
the period of the oscillations does not depend on
$\Gamma$ but their amplitude does.

\newpage

\begin{center}
{\Large {\bf Figure Captions}}
\end{center}

\medskip

Fig. 1 \   The normalized velocity autocorrelation function
$Z(t)$ defined in Eq.\ (\ref{e5}) for $\Gamma=20$ and
magnetic fields  
$B^\ast$=0.0 T. and $B^\ast$= 1.0 ,2.0 , 5.0 , 10.0 and 15.0 T.
The time is measured in units of $\tau$.

\medskip

Fig. 2 \   The velocity autocorrelation function $Z(t)$ for $ \Gamma= $
$20,\ 36,\ 50,\ 90$ and $130$ and magnetic field  $B^\ast$= 1 T.

\medskip

Fig. 3 \    Plot of root mean square displacement
for $\Gamma=20$ and
magnetic fields $B^\ast$= 0  and $B^\ast$=1, 2, 5, 10 and 15 T.

\medskip

Fig. 4 \    Plot of the  self-correlation function $F_s{(q,t)}$
 for five values of $q^*$ for $\Gamma=20$ and $B^\ast$=1.0.

\newpage

\begin{figure}[tbp] 

  \includegraphics[width=8in]{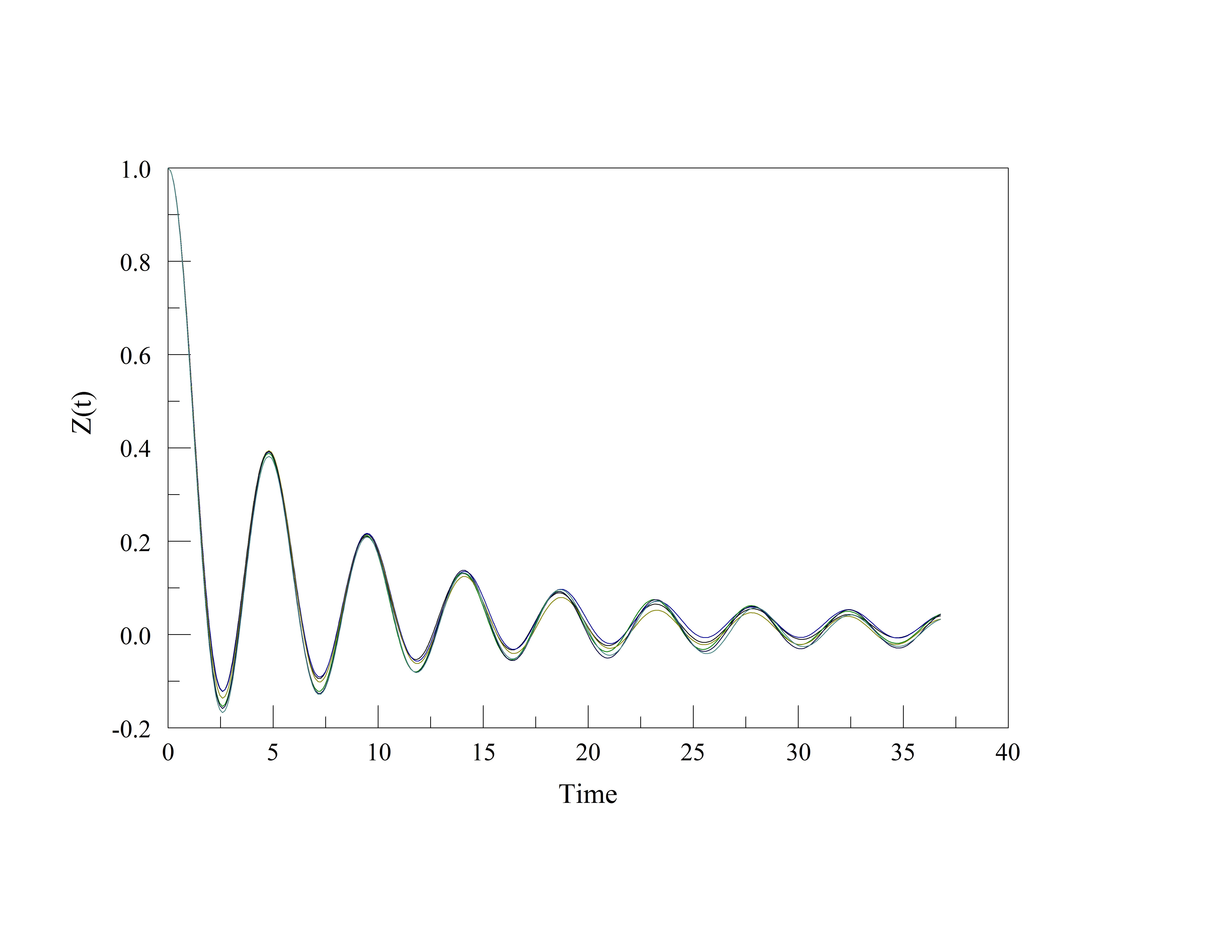}
  \caption{ }
  \label{ }
\end{figure}

\begin{figure}[tbp] 

  \includegraphics[width=8in]{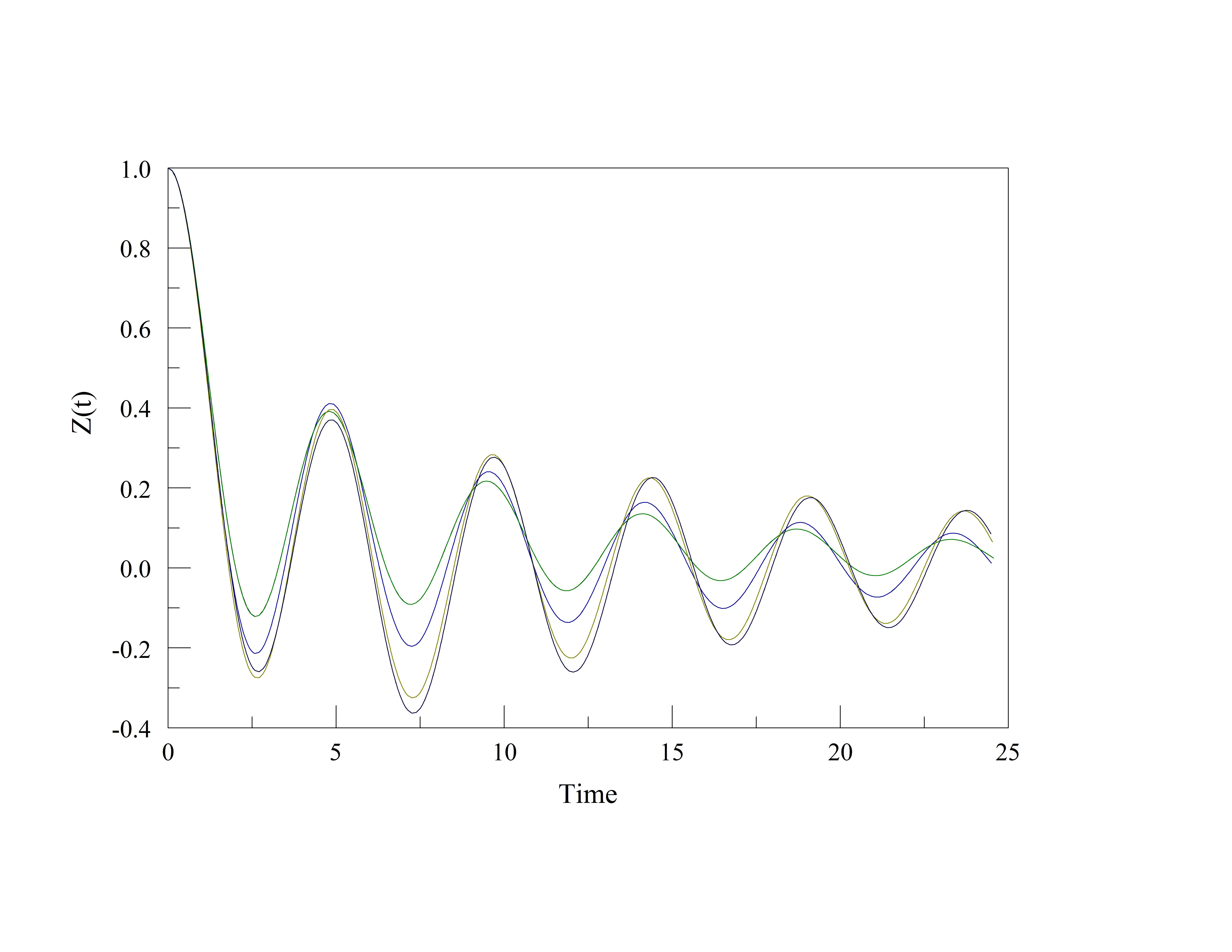}
  \caption{ }
  \label{ }
\end{figure}

\begin{figure}[tbp] 
  \centering
  \includegraphics[width=8in]{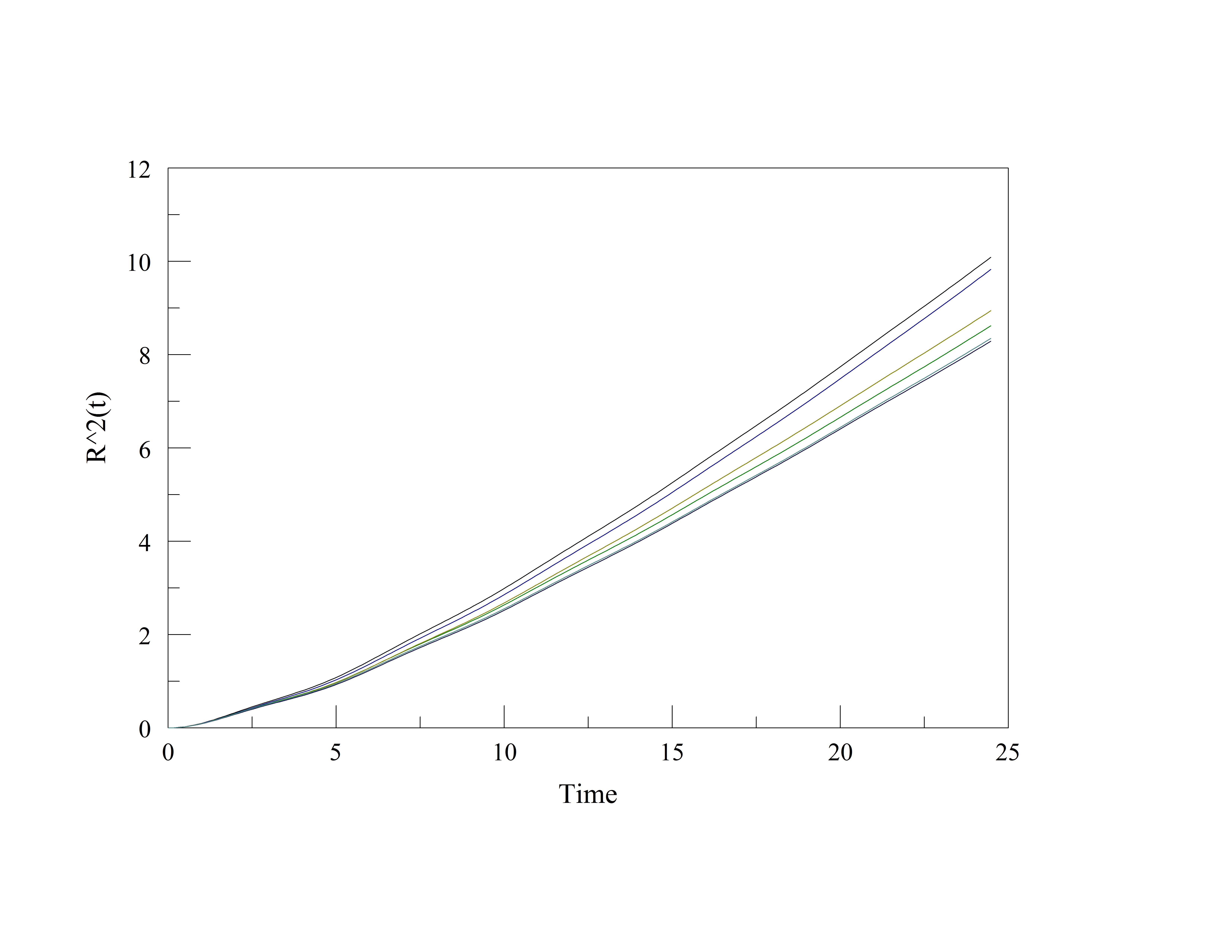}
  \caption{ }
  \label{ }
\end{figure}

\begin{figure}[tbp] 
  \centering
  \includegraphics[width=8in]{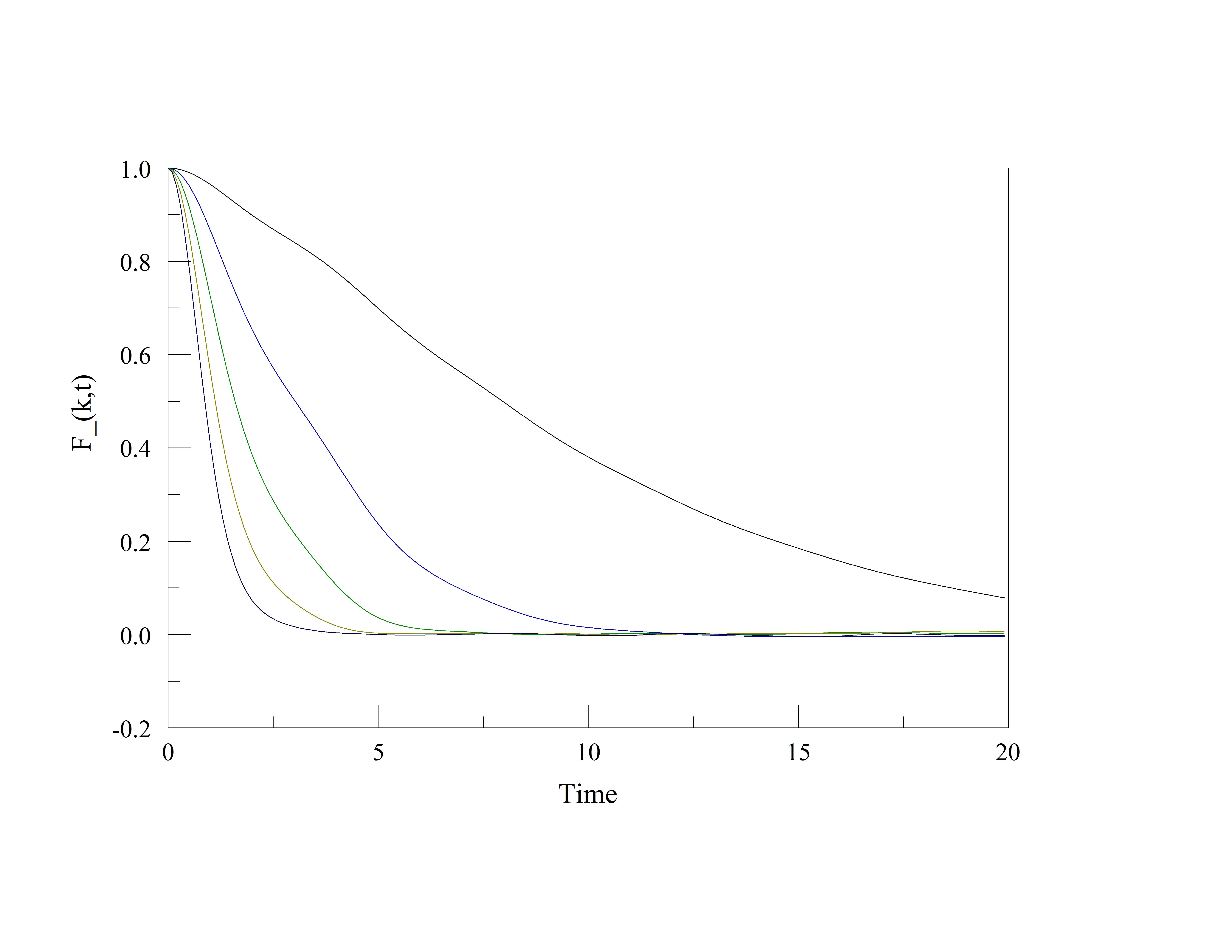}
  \caption{ }
  \label{}
\end{figure}


\begin{thebibliography}{99}



\bibitem{RC2015}  Fabio Deelan Cunden, Anna Maltsev, and Francesco Mezzadri
Phys. Rev. E  {\bf 91}, 060105(R) (2015).



\bibitem{GOD1} T.  Ando,  A. B. Fowler, and F. Stern,  Rev.  Mod.
Phys. {\bf 54},  437 (1982) .

\bibitem{GOD2} J. M. Caillol, D. Levesque, J. J. Weis, and J. P. Hansen,
J. of Stat. Phys., {\bf 28}, 325 (1982).


\bibitem{GOD3}  J. P. Hansen, D. J. Levesque and J. J. Weiss,
Phys. Rev. Letts. {\bf 43}, ,979  (1970).





J. P. Hansen ``Molecular-Dynamics Simulation of
Coulomb Systems in Two and Three Dimensions" xxxxxx



\bibitem{KG1} G. K. Agarwal and K. N. Pathak, J. Phys. C, {\bf  16}, 1887 (1983).


\bibitem{s1}  For a review,  see ``{\em 2D Electron Systems on Helium and
Other Substrates\/}," edited by Y. Andrei (Kluwer Academic, New York, 1997).


\bibitem{s2} K. W. Chiu and J. J. Quinn, Phys. Rev. B {\bf 9}, 4724 (1974).

\bibitem{s3} H. Fuikuyama, Solid State Commun.  {\bf 17}, 1323 (1975).

\bibitem{s4} Lynn Bonsall  and A.A. Maradudin, Phys. Rev. B{\bf 15}, 1959 (1977).

\bibitem{s5} R. Cote and A. H. MacDonald, Phys. Rev. B {\bf 44}, 8759 (1991).


\bibitem{s6} C. C. Grimes and G. Adams, Phys. Rev. Lett. {\bf 42}, 795 (1979).

\bibitem{s7} A. J. Dahn and W. F. Vinen, Phys. Today {\bf 40}, 43 (1988).


\bibitem{s8} Kenneth I. Golden, G. Kalman, and  Philippe Wyns,
Phys. Rev. B {\bf 48}, 8882 (1993).


\bibitem{rr}   N. March and M.P. Tosi,  {\em Coulomb Liquid\/} (Academic
Press, NY, 1992).

\bibitem{rr01}M. W. C. Dharma-wardana and F. Perrot, Phys. Rev. Lett.
{\bf 84}, 959 (2000).

\bibitem{rr02}A. Kallio, P. Pollari, J. Kinaret,  M. Puoskari, in ``
{\em Correlations in the fractional Hall effect}",  Condensed Matter Theories:
Proceedings of the Ninth International Workshop, Edited by F.B. Malik,
(Plenum, New York, NY, USA, 1986) p. 235.

\bibitem{rr1} Toshishige Yamada and D. K. Ferry, Phys. Rev. B{\bf 47}, 6416 (1993).

\bibitem{rr2} J. Cunningham, V. Talyanskii, J. M. Shilton,
 M. Pepper, A. Kristensen, and P. E. Lindelof, Phys. Rev. B {\bf 62}, 1564 (2000).

\bibitem{Wright}  S. J. Wright, M. D. Blumenthal, Godfrey Gumbs, A. L. Thorn,
 M. Pepper, T. J. B. M. Janssen, S. N. Holmes, D. Anderson, G. A. C. Jones,
C. A. Nicoll, and D. A. Ritchie,  Phys. Rev. B {\bf 78}, 233311 (2008).

\bibitem{Graphene1}  Godfrey Gumbs, Andrii Iurov, and  N. J. M. Horing:,
  Phys. Rev.  B  {\bf 91}, 235416 (2015).

\bibitem{Graphene2}   N.  J. M. Horin,  A. Iurov,   Godfrey  Gumbs,
  A. Politano,  and G. Chiarello:
 {\em     Recent Progress in  Nonlocal Graphene Plasmons\/} in
{\bf Low-Dimensional and Nanostructured Materials and Devices}, Springer, pages 205-237  (2015).

\bibitem{Berne}   B. Berne, J. Physiques- Letters {\bf 42}, 251  (1981).


\bibitem{DZH}  K. N. Dzhumagulova, R. U. Masheeva, T. S. Ramazanov, and Z. Donko,
Phys. Rev. E {\bf 89},033104 (2014).




\bibitem{Kalia}  R.  K. Kalia, P. Vashishta,  S. W.  de Leeuw,
 and A. Rahman,  J. Phys C.  {\bf 14}, L991 (1981).


\bibitem{DEL2}  S. W.  de Leeuw, J. W. Perram,  E. R. Smith,
Physica A {\bf  119}, 441 (1983).


\bibitem{agarwal}G. K. Agarwal and K. N. Pathak, J. Phys. C: Solid State
Phys. {\bf 16}, 1887 (1983).


\bibitem{rGGG} J. P. Hansen, I. R. McDonald, and E. L. Pollock,
Phys. Rev. A {\bf 11}, 1025 (1975).

















\end{thebibliography}
\end{document}